# Toward IoT enabled smart offices: Achieving Sustainable Development Goals

**Syeda Nishat Tasnim**

American international University(AIUB)

**Dr. Md Taimur Ahad**

Assistant Professor
American international University(AIUB)

**Abstract:** Despite research advocating the Internet of Things (IoT) as an effective in-office monitoring system, little research has presented societal and climate centric discussions. Whereas the United Nations (UN) and other development agencies concerned with climate impact, are advocating transformative actions towards smart cities, very little research in the IoT domain analyzes the advantages of IoT in achieving sustainable development goals (SDGs) to fill this gap. In this study, a smart office (SO) was developed in a Cisco packet tracer. We then presented the SO through the lens of SDGs. We suggest that SOs support targets mentioned in Goal 6, 7, 8, 9, 11 and 12 of the SDGs. This research is crucial - both for developing and developed economies, as we move toward industrialization, while ignoring the adverse impacts of industrialization. This work is expected to provide a pathway with technological innovation toward a more sustainable world for IT practitioners, governments and development agencies.

## 1. Introduction

The Internet of Things (IoT) refers to the billions of physical objects that are equipped with sensors and use the internet to collect and share data. This technology provides possibilities for transportation, agriculture, healthcare, industrial automation, and emergency responses to natural and man-made disasters [1]. However, whereas most studies concentrate on technical discussion, very few studies discuss the societal perspectives of IoT.

Intended to challenge several adverse social and global challenges, the United Nations (UN) declared a set of Sustainable Development Goals (SDGs). The SDGs aim to end global poverty, protect the environment and nature, and ensure global peace and prosperity. The inter, multi and trans-disciplinary goals offer a path to overcome most overarching problems existing in the current world. Thus, the goals lie at the intersection between nature, society, sustainability and technological transformation.

A careful review suggests that technology has a direct or indirect link to the SDGs. Earlier Kates et al., 2011, also advocated sustainability research is linked to science, and as its' main challenge is to integrate knowledge and methods from different disciplines. It is acknowledged that the most urgent problems sustainability science needs to solve, should be defined by society, not by scientists, thus engagement of the stakeholders in such a process is a condition for success, but also a major challenge.

*Goal 7 Affordable and Clean Energy*, goal 9 *Industry, Innovation, and Infrastructure*, goal 11 *Sustainable Cities and Communities,* and lastly *goal 12 Responsible Consumption and Production* have been examined under the lens of IoT enabled smart office.

The Smart Office (SO) is considered to be a work environment outfitted with IoT devices and therefore linked to the web; it is a smart system made up of multiple connected systems analyzing, regulating, and administering a variety of processes and working circumstances. Observable benefits of SOs, are the utilization of technology to boost productivity and collaboration, automation impacting the organizational environment through reduced energy use, increased safety and simplified operations. However, observable benefits of SO has attracted several authors to shifting to IoT based SO for achieving the SDGs.

In Europe, around 20% of total energy consumed is contributed by non-residential buildings; globally, this consumption has continued to increase. Before 2020, smart building proposals (focused on Nearly Zero Energy Buildings (NZEB), air quality monitoring, and energy-saving with thermal comfort etc.) were necessary, but due to the pandemic this area of R&D has become more essential. Moreover, the necessity to unite the SDGs and gain technological solutions centered on IoT, involves holistic contributions through actual installations that serve as spaces for measurement and testing. Despite the significant importance of IoT, limitations such as information security and assurance, and information quality are also raised by authors [13]. Realizing such limitations, we present a model of SO which comprises two primary parts, including the server (webserver) which provides a framework controlling and screening clients' homes. The system code can be controlled locally (LAN) or from a distance (web). The second component is the equipment interface module, which provides a suitable interface to sensors and actuators of the home computerization framework.

## 2. Literature Review

Two perspectives of works of literature are considered in this study, firstly, IoT studies on SO and secondly IoT studies attempting to achieve SDGs. In the case of the former literature, Ajao et. al. applied android based web apps in Wi-Fi technology for an embedded system based IoT for smart home and office appliances. For easy control and operation of domestic home appliances on a remote network, the internet protocol (IP) addresses are assigned to devices, which aid interoperability and end-to-end communication. However, the paper suggested a wireless network between the home server and home devices using low power consumption wireless technology should work in the future [2]. El Shafee and Hamed also had the same idea towards applying the IoT based Smart home [4].

Emeyazia et. al. designed and developed an Office Automation System based on the IoT and Voice Recognition Command. The research aimed to develop a system that can control the appliances with voice commands. Arduino IDE framework is used for the development of the system, C++ programming language is used for the programming of the system and This Then That (IFTTT) platform has been used to build a voice control. The system was able to control the appliances by the voice command with the usage of Google home mini device. The laboratory test result showed that the system has a high (about 98%) performance rating. However, the system is still operated with the human command, thus the appliances like printer, air conditioner, light, fan, etc. will continue to be switched on unless and until there is a command to switch off and simultaneously can also cause mode error. [3]

Such et. al. presented a plan of intelligent office frameworks for controlling, checking and mechanization electrical appliances depending on the worker's entry and exit in the work area. The information is shown in the screen by using the data in the cloud about the entry and exit timings, several cabins and power utilization of the workers utilizing the RFID labels filters at the passage point of the workplace, it additionally permits and helps the workers to control and screen their cabin's electricity status through a cell phone; this framework helps reduce the pressure of work carried out by workers. [6]

Martínez et. al. implemented an IoT ecosystem at a university. The installation showed effectiveness by focusing on $CO_2$ and energy consumption monitoring, demonstrating the IoT potential as SDG-enabling technologies. These contributions not only allow experimental lab tests (from the authors' expertise in several specialties of Industrial, Mechanical, Design, Thermal, Electrical, Electronic, Computer and Telecommunication Engineering), but also a reference model potentially used in direct application to research projects and institutional initiatives, academic research extendable to professional environments, buildings and cities. [13]

Nastic et al. applied middleware to offer comprehensive support for multi-level provisioning of IoT Cloud systems. The features of the middleware such as generic, lightweight resource abstraction and mechanisms allowing application-specific customization of IoT devices to support IoT resources and application components. The study's implementation of middleware has been evaluated through the use of real-life applications in building management systems. [15]

Dalal proposed the potential role of IoT in accomplishing the target of sustainable development worldwide. The study was performed as an attempt to find gaps, challenges and obstacles in reaching SDG targets. Their findings showed the significance of IoT in achieving the SDG, in which 84% of respondents believed accomplishing sustainable development and growth of the green economy, through Green IoT and Green ICT technologies can act as a game-changer. The areas where technology can act as a major asset, include productive and efficient use of raw materials, preserving natural resources and decreasing emissions of greenhouse gases and other waste. A globally sustainable environment requires the planned and organized use of IoT and G-IoT concepts for constructing or developing goods and services for enriching renewable energy sources, energy-conserving computing, management of power sources, green metrics. Assessment tools and methodologies have substantial potential to transform traditional development into sustainable development. [16]

## 3. Research Methodology

A qualitative methodology was adopted here, as it is suitable for problem-centric research. In computing studies and the IT discipline in general, there is an increase in the use of qualitative research methods, possibly because qualitative research offers several approaches in dealing with problem-centric research, for example the experimental method. An experimental approach has been applied here. An experiment is performed to discover a meaningful solution to a problem. The experiment must contain the answer to its question. Beveridge noted "an experiment usually entails causing an occurrence to occur under controlled settings in which as many external influences as possible are excluded and close observation is enabled to reveal relationships between phenomena" (p. ).

## 4. Details of the experiment

Figure 1 depicts the concept of SO as used here. The diagram shows how sensor data is handled in communication between the mobile phone and the sensor.

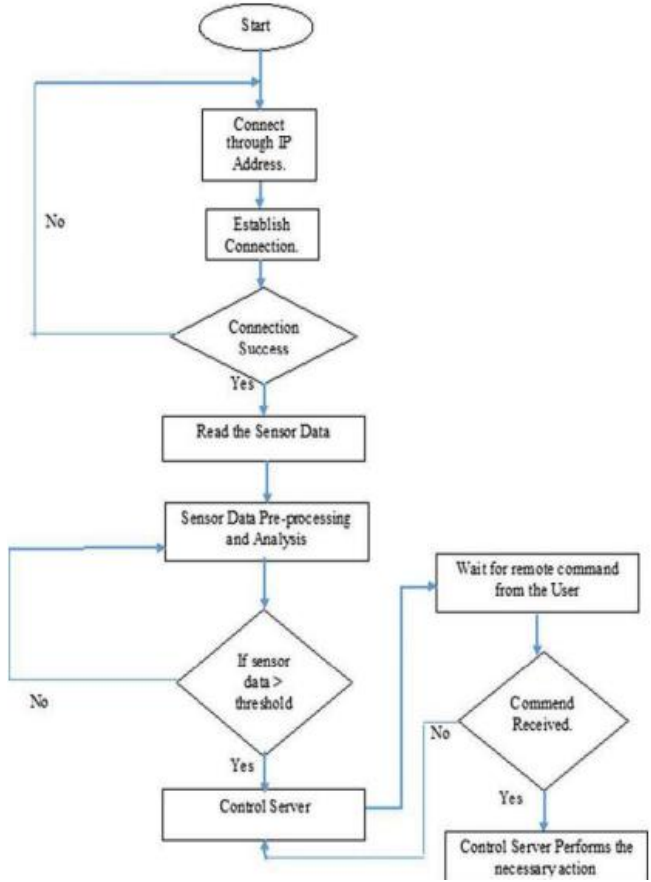

Figure 1: Flow Chart

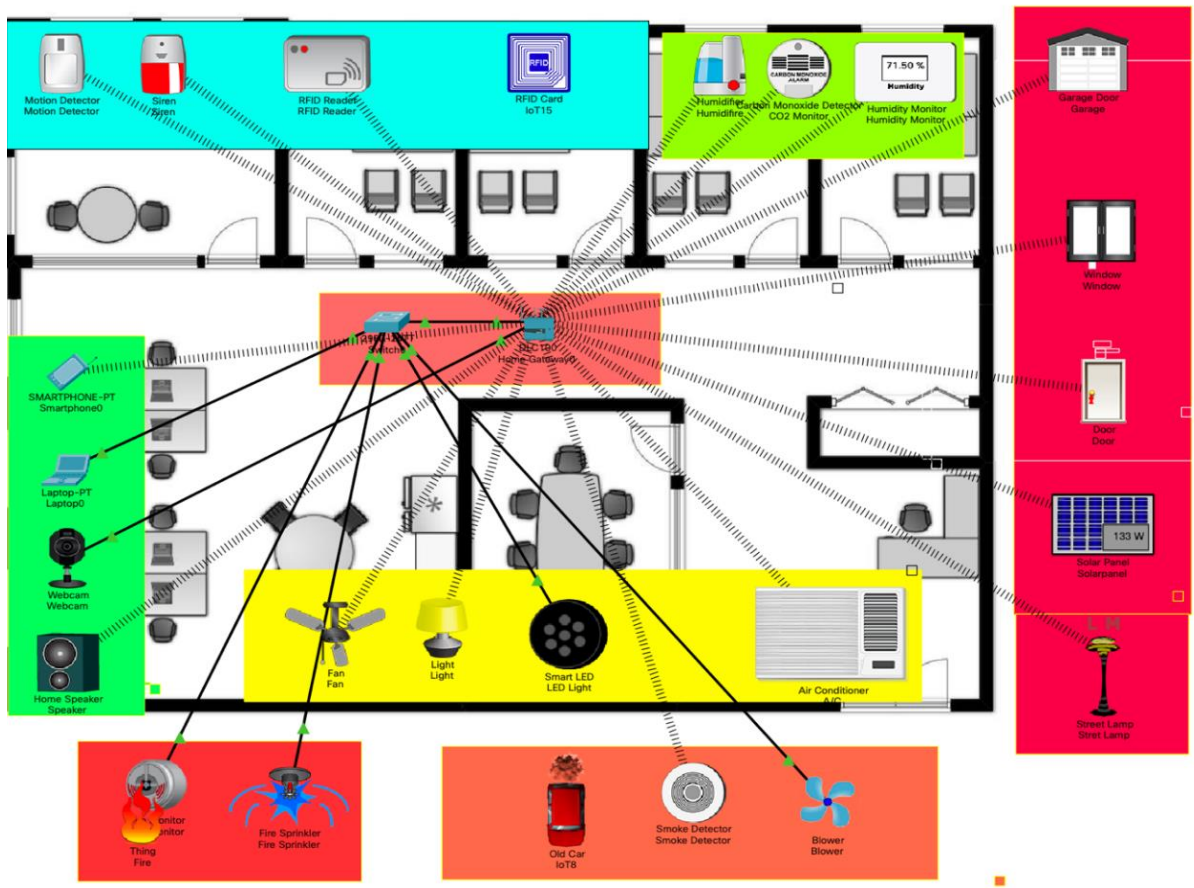

Figure 2: The design view

To visualize the concept, *Cisco Packet Tracer* was used. A *Cisco Packet Tracer* offers IoT tools such as motion detector, siren, RFID reader, RFID Card IoT15, SMARTPHONE-PT, webcam, Light, Home Speaker, Fire Monitor, Fire Sprinkler, Fire Sprinkler, Humidifier, Carbon Monoxide Detector, Humidity Monitor, Switch and office Gateway, which were essential to design the SO.

Figure 3: Applied devices

## 4.1 Implementation

The steps of connecting IoT devices are described below:

i. IoT devices, routers, switches, smartphones, servers, and wires are physically placed.
ii. All routers/switches are connected. IoT devices/home devices/servers & smartphone are connected to the routers/switches.
iii. Router 0/ Router 1 is configured using the DHCP protocol to provide IP addresses.
iv. Routing protocol EIGRP is configured in router 0 to access the SO by using a smartphone.
v. An account is created to access the system.

Figure 4: Control view

## 4.2 Conditions applied

i. A motion detector and webcam detect people standing at the front of the office (figure 5).
ii. In case the lawn sprinkler is turned on, the water drain will start draining.
iii. In case a wind detector detects high wind, then all windows of the home will be closed.
iv. In the office is occupied, the fan, light and window will be turned on or opened.
v. In case the air conditioner is turned on, the windows of the office will be closed.
vi. In the car or anyone is in the front of the garage, then the motion detector will close the door.

| Actions | Enabled | Name | Condition | Actions |
|---|---|---|---|---|
| Edit Remove | Yes | Motion On | Motion Detector On is true | Set Webcam On to true / Set Siren On to true |
| Edit Remove | Yes | Smoke On | Window On is true | Set Siren On to true |
| Edit Remove | Yes | Motion Off | Motion Detector On is false | Set Webcam On to false / Set Siren On to false |
| Edit Remove | Yes | Smoke Off | Window On is false | Set Siren On to false |
| Edit Remove | Yes | Sprinkler On | Fire Monitor Fire Detected is true | Set Fire Sprinkler Status to true / Set Siren On to true |
| Edit Remove | Yes | Sprinkler Off | Fire Monitor Fire Detected is false | Set Fire Sprinkler Status to false / Set Siren On to false |
| Edit Remove | Yes | Smoke On car | Smoke Detector Level >= 0.18 | Set Blower Status to High / Set Window On to true |
| Edit Remove | Yes | Smoke car off | Smoke Detector Level < 0.1 | Set Blower Status to Off / Set Window On to false |
| Edit Remove | Yes | RFID Valid | RFID Reader Card ID = 1001 | Set RFID Reader Status to Valid |
| Edit Remove | Yes | RFID invalid | RFID Reader Card ID != 1001 | Set RFID Reader Status to Invalid |
| Edit Remove | Yes | Door Unlock | RFID Reader Status is Valid | Set Door Lock to Unlock |
| Edit Remove | Yes | Door Lock | RFID Reader Status is Invalid | Set Door Lock to Lock |

Figure 5: Applied conditions

## 5. Discussion

A careful review suggests that technology and science have a direct or indirect link to the SDGs. Earlier Kates et al., 2011, also advocated sustainability research is linked to science, and as its' main challenge is to integrate knowledge and methods from different disciplines, as the goals are either inter, multi and trans-disciplinary; this requires a stakeholder-oriented approach and methodological innovation (Schoolman et al., 2011). It is acknowledged the most urgent problems that sustainability science needs to solve should be defined by society, not by scientists, thus engagement of stakeholders in such a process is a condition for success, but also a major challenge. However, goal 7-Affordable and Clean Energy, goal 9-Industry, Innovation, and Infrastructure goal 11-Sustainable Cities and Communities, lastly goal 12-Responsible Consumption and Production have been subjects of study here, through the lens of an IoT enabled smart office (figure 6). However, the SO presented, is expected to provide to following SDGs:

6.4: increase water-use efficiency across all sectors

7.1: ensure universal access to affordable, reliable and modern energy services

7.3: double the global rate of improvement in energy efficiency.

7b: Upgrade technology for sustainable energy

8.4: global resource efficiency in consumption and production

9.4: greater adoption of clean and environmentally sound technologies

9.5: upgrade the technological capabilities of industrial sectors

11.6: reduce the adverse impact of cities

12.2 sustainable management and efficient use of natural resources.

12.5: substantially reduce waste generation

13. b: Promote climate change-related planning and management

## 6. Conclusion

The purpose of this research was to present an idea of SO and relate this with achievable SDGs goals. Therefore, the work presented here is only narrative, as experimental results are not qualitatively or quantitatively tested. Another limitation is that the study is also not empirical; rather a concept simulated by a packet tracer. However, one defense is that the observable benefits of IoT success in SO, addresses the need for the development of SOs by providing a conceptual introduction to IoT based SO, as a practical guide for researchers and industry practitioners. Future works and aims are therefore to implement a real-life scenario and examine outcomes using qualitative or quantitative judgement. Climate-related adverse and supportive particulars should be a subject of study using IoT based sensors. Artificial intelligence coupled with IoT may address the SDGs goals. Limitations of IoT devices and communication in SO can be also a subject for further study. However, to our knowledge, this is the first research attempt in shading lights of IoT based SO in achieving SDGs. The research will provide subsequently value to both developed and developing nations for a sustainable world in the coming future.

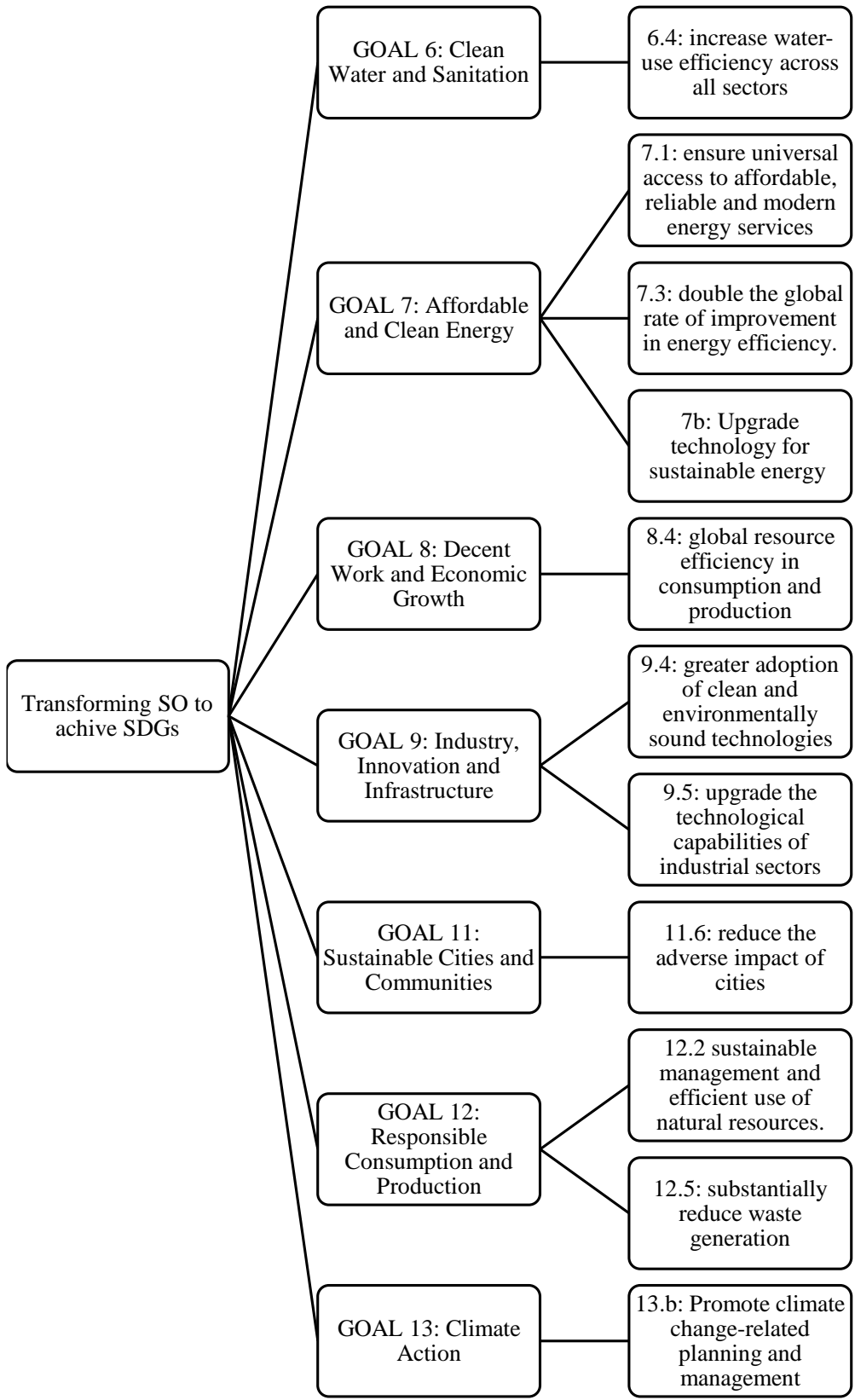

Figure 6: Relevant SDG goals